\pdfoutput=1
\documentclass[a4paper,12pt]{article}
\usepackage{amsfonts}
\usepackage{mathrsfs}
\usepackage{amsmath}
\usepackage{amssymb}
\usepackage{framed}
\usepackage[medium]{titlesec}
\usepackage{bm}
\usepackage{cite}
\usepackage[normalem]{ulem}
\usepackage{extarrows}
\usepackage{slashed}
\usepackage{isodateo}
\usepackage{graphicx}
\usepackage{xcolor}
\usepackage[bookmarksnumbered=true,bookmarksopen=true]{hyperref}
 \hypersetup{colorlinks,%
             linkcolor=[rgb]{0,0.3,0.6}, %
             citecolor=[rgb]{0,0.3,0.6}, %
             urlcolor=[rgb]{0,0.3,0.6}}
\usepackage[hmargin=.7in,vmargin=1.1in]{geometry}
\usepackage{indentfirst}
\newcommand{\FR}[2]{\displaystyle\frac{\,{#1}\,}{#2}}

\renewcommand{\rm}{\mathrm}

\graphicspath{{fig/}}

\def\bge{\begin{equation}}
\def\ede{\end{equation}}
\def\bga{\begin{aligned}}
\def\eda{\end{aligned}}
\def\bgp{\begin{pmatrix}}
\def\edp{\end{pmatrix}}
\def\bgs{\begin{subequations}}
\def\eds{\end{subequations}}
\newcommand{\order}[1]{\mathcal{O}({#1})}
\def\di{{\mathrm{d}}}
\def\D{{\mathrm{D}}}
\def\Di{{\mathcal{D}}}

\def\mb{\mathbf}

\def\pd{\partial}
\def\ld{{\mathcal{L}}}

\def\la{\langle}\def\ra{\rangle}

\def\to{\rightarrow}
\def\To{\Rightarrow}
\def\ii{\mathrm{i}}

\def\al{\alpha}
\def\be{\beta}
\def\ga{\gamma}
\def\de{\delta}
\def\ep{\epsilon}
\def\ka{\kappa}
\def\lam{\lambda}
\def\rh{\rho}
\def\si{\sigma}

\newcommand{\ob}[1]{\mkern 2mu \overline{\mkern -2mu #1 \mkern -2mu}\mkern 2mu}
\newcommand{\wt}[1]{\mkern 2mu \widetilde{\mkern -2mu #1 \mkern -2mu}\mkern 2mu}

\newcommand{\phidot}{{\dot \phi_0}}

\newcommand{\fo}{f_\text{NL}^\text{(osc)}}
\newcommand{\nuf}{\nu_{{1}/{2}}}
\newcommand{\nuv}{\nu_{1}}

\begin{document}

\title{\Large\textbf{In Search of Large Signals at the Cosmological Collider}}
\author{Lian-Tao Wang$^a$\footnote{Email: liantaow@uchicago.edu}~~~and~~~Zhong-Zhi Xianyu$^b$\footnote{Email: zxianyu@g.harvard.edu}\\[2mm]
\normalsize{\emph{$^a$~Department of Physics, University of Chicago, Chicago, IL 60637}}\\
\normalsize{\emph{$^b$~Department of Physics, Harvard University, 17 Oxford Street, Cambridge, MA 02138}}}

\date{}
\maketitle

\begin{abstract}
  
We look for oscillating signals in the primordial bispectrum from new physics heavy particles which are visibly large for next generation large scale structures (LSS) survey. We show that in ordinary inflation scenarios where a slow-rolling inflaton generates density fluctuations and with no breaking of scale invariance or spacetime symmetry, there exist no naturally large signals unless the rolling inflaton generates a parity-odd chemical potential for the heavy particles.  We estimate the accessibility of this signal through observations. While current CMB data are already sensitive in the most optimistic scenario, future probes, including LSS survey and 21 cm observation, can cover interesting regions of the model space.
\end{abstract}

\section{Introduction}

Heavy particles from new physics can be produced on-shell during cosmic inflaton. It has been emphasized recently that these particles can then impact on spacetime fluctuations and leave unique imprints on the cosmic microwave background (CMB) and large scale structures (LSS) \cite{Chen:2009zp,Chen:2012ge,Arkani-Hamed:2015bza}. Specifically, in the squeezed limit of the 3-point correlation of the curvature fluctuations where the wave number of one mode is much smaller than the other two, the on-shell heavy particles can generate distinct shape dependence. This includes a nonanalytic oscillatory or scaling behavior as a function of momentum ratio, and  a particular dependence on the angle between the long mode and the short mode. This gives us the opportunity to search for new physics at an energy scale far beyond the reach of any foreseeable terrestrial experiments. In addition, the shape dependence carries the information about the mass and the spin of the new particles \cite{Lee:2016vti,Chen:2016nrs,Chen:2016uwp,Chen:2016hrz,Chen:2017ryl,Chen:2018xck,Chen:2018sce,An:2017hlx,Iyer:2017qzw,Kumar:2017ecc,Kumar:2018jxz,Kumar:2019ebj,Wang:2018tbf,Lu:2019tjj,Hook:2019zxa,Hook:2019vcn}.

The oscillatory shape dependence induced from the heavy particle is quite unambiguous. It is a direct consequence of the inflating background and the on-shell particle production, and is largely independent of the details of the inflation models and the inflaton-matter couplings. On the other hand, the size of the signal is very sensitive to the details of the coupling. It is often the case that the oscillatory signal is too small to be observed in any future probes of the 3-point correlation, commonly known as the non-Gaussianity (NG). It is thus important to ask which process could generate large enough signals to be searched for in future probes.

The size of NG is commonly quoted by a dimensionless parameter $f_\text{NL}$. The recent CMB constraints from Planck gives that $f_\text{NL}\lesssim \mathcal{O}(10)$, with mild shape dependence \cite{Akrami:2019izv}. In the coming decade, the LSS surveys such as SPHEREx \cite{Dore:2014cca} could measure local shape to roughly $\mathcal{O}(1)$. More futuristically, it has been show that 21cm tomography could probe $f_\text{NL}\sim\order{0.01}$, eventually reaching the gravitation floor\footnote{Here by gravitational floor we mean the NG generated by pure gravitational interactions among inflaton fluctuations. In the typical slow-roll inflation this is on the order of slow-roll parameter $\sim\order{0.01}$. The gravitation floor itself makes an interesting target for future observation. On the other hand, we expect that the sensitivity to the signals studied in this paper would be better thanks to its distinct shape.} of non-Gaussinity \cite{Munoz:2015eqa,Meerburg:2016zdz}. See also \cite{Meerburg:2019qqi}. With these observational targets in mind, it is useful to look for scenarios with oscillatory signals $f_\text{NL}^\text{(osc)}\gtrsim\order{1}$ that can serve as targets of LSS survey of near future. More optimistically, it is also useful to identify scenarios with $f_\text{NL}^\text{(osc)}>\order{0.01}$ which is within the reaches of the future probes. Of course, the oscillatory behavior is only a part of the non-Gaussian signal and the oscillation amplitude $f_\text{NL}^\text{(osc)}$ can not be directly compared with the projected reach of general NG directly. On the other hand, a $f_\text{NL}^\text{(osc)}$ which is too much below it would not be observable.    

In this paper, we consider the simplest inflation scenarios with single field slow roll background, and classify couplings between inflaton and matter fields according to whether they can generate large oscillatory signals. We parameterize the couplings between the inflaton and matter fields with effective field theory (EFT) operators up to dimension-6 (dim-6). We impose following conditions on the inflaton-matter couplings: 1) dimensionless couplings being $\order{1}$; 2) a single cut-off scale consistent with EFT expansion; 3) no large cancellation among different terms in the Lagrangian (no fine tuning) at the tree level.

Under these conditions, surprisingly or not, almost all couplings fail to generate visibly large signals. The difficulty stems from the fact that a large signal would need large coupling, which would in turn introduce large mass corrections to the matter field due to its coupling with the inflaton background ($\phi_0$ or $\dot\phi_0$). This tends to make the matter field too heavy to be excited during inflation. There is only one exception to this nearly no-go result, namely the dim-5 operators generating chemical potential to matter fields through the rolling inflaton background. There is one further exception, the well-studied quasi single field inflation \cite{Chen:2009zp}, if we give up the first condition and allow for a tiny dimensionless coupling which we will quantify in the discussion. Without question, these exceptional cases deserve more detailed study.

The scenario of quasi single field inflation has been extensively studied in the context of non-Gaussianities. The chemical potentials are relatively less known, but were also applied recently in several studies of cosmological collider physics. Given its uniqueness in generating large signals, a more comprehensive analysis is clearly needed. This will be the second theme of this paper.

Naively, one may expect that any type of chemical potential from the rolling inflaton could help to enhance the particle production and generate large signals. However, we will show that, without further breaking the symmetries (in particular the spatial rotations and the scale invariance), the only viable case is the chemical potentials associated with a parity-odd conserved charge. We provide intuitive arguments for this conclusion and justify it with explicit calculations. 

We should emphasize that we are considering the most economic and conservative scenarios. 
It will also be useful to consider more general possibilities. Known examples of large signals in non-minimal scenarios include resonant production of heavy particles from broken scale invariance \cite{Flauger:2016idt}, generating the curvature fluctuation by a spectator field via either modulated reheating \cite{Lu:2019tjj} or curvaton mechanism \cite{Kumar:2019ebj}, breaking background symmetry by a gauge field vacuum expectation value (VEV) \cite{Chua:2018dqh} or by warm inflation \cite{Tong:2018tqf}. 

\begin{figure}[tbph]
\centering
\includegraphics[width=0.8\textwidth]{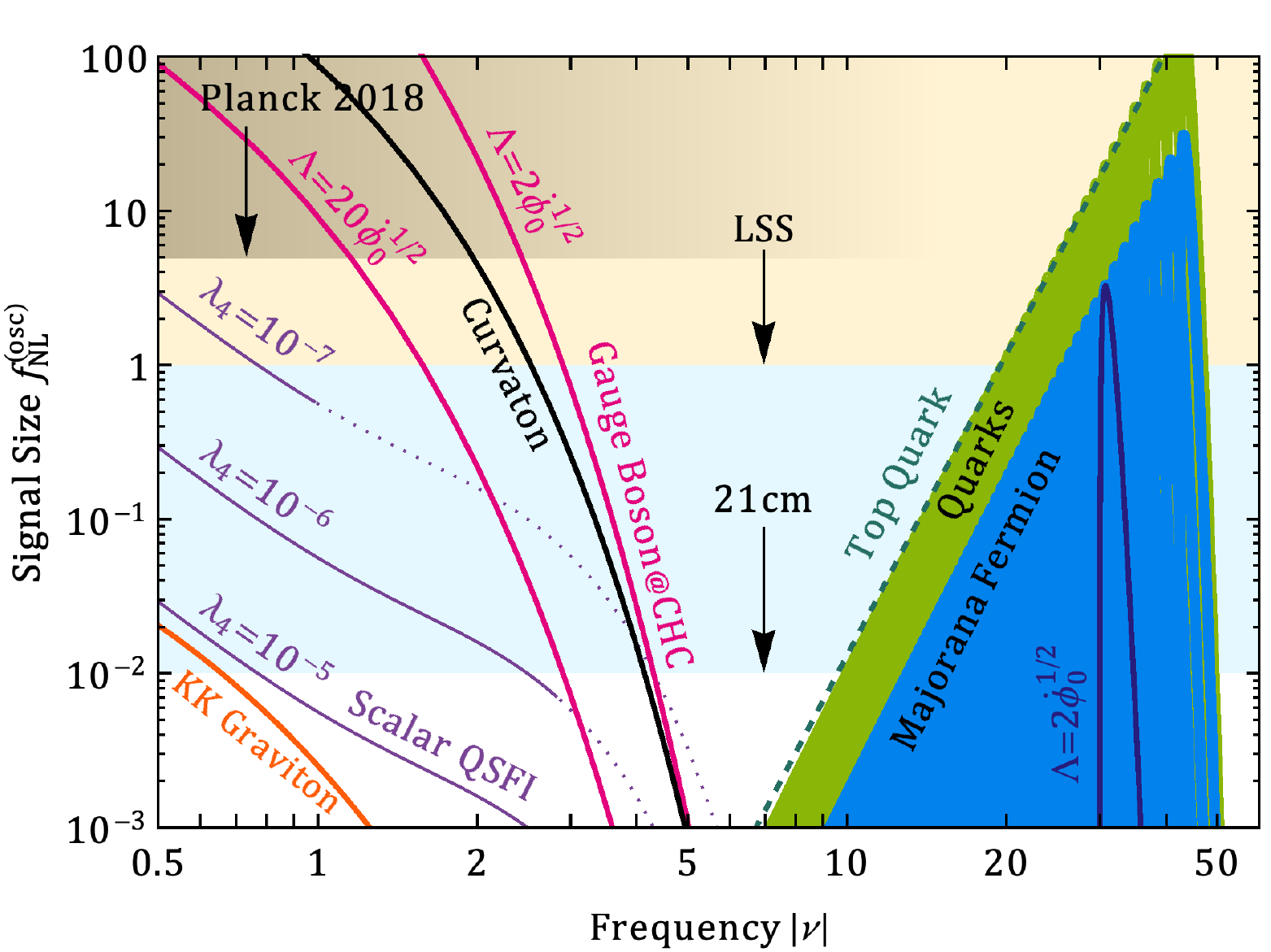}
\caption{The signal strength $f_\text{NL}^\text{(osc)}$ of oscillations in the squeezed bispecetrum as functions of the frequency $|\nu|$ in various scenarios. The frequency depends on the relative size of the mass of the particle being produced, with $\nu = \sqrt{9/4 - (m/H)^2}$ for scalar and $\nuv = \sqrt{1/4-(m/H)^2}$ for gauge boson. For fermion production, it has an important dependence on the chemical potential $\mu=\dot \phi_0/\Lambda$, with $\nuf = (\sqrt{m^2+\mu^2})/H$, and for the top condensate $m$ and $\mu$ are dynamically related by $m=\sqrt{\mu H/\pi}$ \cite{Hook:2019vcn}. $\Lambda$ is the cut off scale of the EFT used to characterize the coupling between the inflaton (Higgs boson for the CHC) and the particle being produced.  The shaded wedges are from fermion productions \cite{Chen:2018xck,Hook:2019zxa,Hook:2019vcn}, where we scan the cut off scale from $\Lambda^2 = 2 \dot \phi_0$ to larger values. ``Majorana Fermion" refers to the case of a singlet Weyl fermion. The signal from one flavor of SM quarks is six times larger due to degrees of freedom counting (2 from Dirac and 3 from color). The case of top quark is different since it depends directly on the top quark mass \cite{Hook:2019vcn}.  The predictions from scalar Quasi-Single Field Inflation (QSFI) \cite{Chen:2017ryl,Kumar:2017ecc} as a function of the scalar quartic coupling are represented with purple curves, where the dotted part indicating the strong coupling regime.  We have also included the signal prediction from tree-level Z boson exchange in the cosmological Higgs collider scenario (magenta) \cite{Lu:2019tjj} , tree-level Higgs exchange from the curvaton scenario (black) \cite{Kumar:2019ebj}, and the tree-level exchange of KK-graviton (orange) \cite{Kumar:2018jxz}. For comparison, we have displayed the reach of NG from the current Planck data, the LSS survey, and the 21 cm observations as horizontal shaded regions. See text in Sec.~\ref{sec_discussions} for further explanations and discussions. }
\label{Fig_signal}
\end{figure}

In Fig.~\ref{Fig_signal}, we summarize the signal strengths $f_\text{NL}^\text{(osc)}$ of several known scenarios as functions of the oscillation frequency $|\nu|$, including the chemical-potential-enhanced fermions, a scenario of Cosmological Higgs Collider (CHC), the curvaton scenario, the scalar quasi-single field inflation (QSFI), as well as the KK graviton in extra dimension models. We also show the current limit and future reach from CMB, LSS, and 21cm observations. More information and explanations of the figure are collected in Sec.~\ref{sec_discussions}.

This rest of this paper is organized as follows. In Sec.~\ref{sec_size} we briefly review the cosmological collider signals and estimate the signal strength for various EFT couplings between the inflaton and matter fields in the minimal scenario. We show that, without fine tuning, all couplings up to dim-6 operators lead to tiny signal unless the coupling generates chemical potentials for the heavy particle. We then provide intuitive understanding of the chemical potential in Sec.~\ref{sec_intuitive}, and explain why only parity-odd chemical potential can potentially enhance the signal. We justify this intuitive picture in Sec.~\ref{sec_mode} by explicit mode functions for particles of various spins with nonzero background chemical potential. The result of this section will also be useful for more detailed study of chemical potential scenarios. In Sec.~\ref{sec_size_chempot} we redo the signal estimate for chemical potential scenarios and show that they can naturally lead to large signals. More discussions and a detailed explanation of the Fig.~\ref{Fig_signal} are collected in Sec.~\ref{sec_discussions}.

\section{Size of the Oscillatory Signal}
\label{sec_size}

In this section we look for inflaton-matter couplings that can generate large cosmological collider signals. We will not consider the most general possibilities. Rather, we will restrict ourselves to a minimal scenario where the inflation is driven by a single slowly rolling inflaton field $\phi$, and curvature perturbation $\zeta$ is also generated from the quantum fluctuation $\de\phi$ of the inflaton field in the standard way. We will then consider general EFT couplings between the inflaton and some matter fields up to dim-6 operators. Throughout the section, we will impose the following conditions:
\begin{enumerate}
  \item All dimensionless couplings are of $\order{1}$.
  \item Non-renormalizable couplings are controlled by a single cutoff scale $\Lambda$, which is large enough to justify only keeping the lower order terms in the EFT expansion.
  \item No fine tuning of parameters at the tree level. For example, we require that there should not be large cancellation between different terms in the Lagrangian to the mass of the matter particle. 
\end{enumerate}
These are not requirements based on consistency of the theory, but on naturalness considerations. A natural scenario will be more generic and is in principle easier to be realized in new physics models. Large cancellation can be hard to justify from model building point of view. For example, we could have the mass correction coming from the inflaton background cancel against the original mass of the matter field. However, in general, these two contributions are from unrelated origins, and we would expect that such a cancellation is accidental.

We also note that, from the point of view of a complete model,  no fine tuning at the tree level is not enough. Quantum corrections to the mass of the matter field, in particular scalar matter fields, need to be regulated in any UV completion to avoid fine-tuning. Without going into building such a complete model, we assume that it has been addressed by some mechanism. 
It is interesting to observe that the scenario of chemical potential can still lead to naturally large signal even we exclude loop-level fine tuning, since this mechanism works only for fermions or gauge bosons, whose mass can be made stable against loop corrections.  

There is actually a more general EFT framework for inflation which we could use to perform our analysis \cite{Cheung:2007st}. This EFT is designed to capture phenomenologically the fluctuating modes in the CMB sky and it is agnostic about the rolling inflaton and thus can be applied to very broad inflation scenarios. We do not choose this framework because the naturalness consideration is essentially about the underlying model, which the EFT framework does not capture. It also obscures the derivative expansion in terms of $\dot\phi$, which makes the relative importance of operators less transparent. 

\subsection{Brief review of cosmological collider signals}
 Before moving on to a survey of EFT couplings, we first review very briefly the cosmological collider signals and how to estimate their size. More detailed reviews on this topic can be found in \cite{Chen:2017ryl,Baumann:2018muz}. In this paper, we only consider the simple inflation scenario with a slow-roll inflaton $\phi$ which provides the vacuum energy driving the inflation and also generates the observed density fluctuations through its own quantum fluctuations $\de\phi$ during inflation. There are non-minimal alternative scenarios that can be interesting too \cite{Flauger:2016idt,Tong:2018tqf,Lu:2019tjj,Kumar:2019ebj}. During inflation $\de\phi$ can be treated as a nearly massless field which acquires quantum fluctuations which are nearly Gaussian, with the variance in momentum space given by $\la\de\phi_k\de\phi_{-k}\ra'=2\pi^2k^{-3}P_{\de\phi}$ and $P_{\de\phi}=(H/2\pi)^2$ is the nearly scale invariant power spectrum of the inflaton fluctuation. Here and in the following a prime $\la\cdots\ra'$ means the $\de$-function of momentum conservation being removed. We can convert inflaton fluctuation $\de\phi$ to the curvature fluctuation $\zeta$ through a redefinition of time slices as $\zeta=-(H/\dot\phi)\de\phi$. From this we know that the power spectrum of the curvature fluctuation is
\begin{align}
  P_\zeta =\FR{H^2}{\dot\phi_0^2}\FR{H^2}{(2\pi)^2}.
\end{align}
The nearly scale invariant power spectrum is measured to be $P_\zeta \simeq 2\times 10^{-9}$ at CMB scales. From this we have the relation $\dot\phi_0^{1/2}\simeq 60 H$. 

For the discussion of cosmological collider physics we will mostly focus on the 3-point correlation of the curvature perturbation $\zeta$. In 3-momentum space the 3-point correlation $\la\zeta_{k_1}\zeta_{k_2}\zeta_{k_3}\ra$ is a function of momenta triangle formed by $\mb k_i~(i=1,2,3)$. The dimensionless bispectrum $S(k_1,k_2,k_3)$ is conventionally defined through the above 3-point correlation through
\begin{align}
\la\zeta_{k_1}\zeta_{k_2}\zeta_{k_3}\ra'=(2\pi)^4P_\zeta^2\FR{1}{(k_1k_2k_3)^2}S(k_1,k_2,k_3),
\end{align}
where $k_i\equiv |\mb k_i|$, and the overall amplitude of the bispectrum gives the strength of NG, traditionally denoted by $f_\text{NL}\simeq |S(k_1,k_2,k_3)|$. Of course the size of $S(k_1,k_2,k_3)$ depends on the momentum configuration so one should in principle define $f_\text{NL}$ for different shapes separately. We will mostly focus on the squeezed limit of the bispectrum where one of the three momenta is much smaller than the other two, i.e., $k_3\ll k_1\simeq k_2$. When a massive field is produced on shell and is converted to the soft mode $\de\phi_{k_3}$, it will leave characteristic non-analytical dependence on the momentum ratio $k_3/k_1$. In this paper, by (cosmological collider) signals we will always mean this non-analytical dependence in the squeezed limit and we denote the overall amplitude of the signal by $f_\text{NL}^\text{(osc)}$. 

To estimate the size of the signal, we first convert the $\zeta$-correlations to $\de\phi$-correlations,
\begin{align}
  S(k_1,k_2,k_3)=&-\FR{(k_1k_2k_3)^2}{(2\pi)^4P_\zeta^2}\Big(\FR{H}{\dot\phi_0}\Big)^3\la\de\phi_{k_1}\de\phi_{k_2}\de\phi_{k_3}\ra'.
\end{align}
The $\de\phi$-correlations can then be calculated following Schwinger-Keldysh formalism, which is essentially Feynman-diagram expansion. See \cite{Chen:2017ryl} for a pedagogical review. In this way, the size of the signal, $f_\text{NL}^\text{(osc)}\sim |S|$, can be estimated as,
\begin{align}
\boxed{~~~~
f_\text{NL}^\text{(osc)}\sim\FR{1}{2\pi P_\zeta^{1/2}}\la\de\phi_{k_1}\de\phi_{k_2}\de\phi_{k_3}\ra'
\sim \FR{1}{2\pi P_\zeta^{1/2}}
\times\text{loop factor}
\times\text{vertices}\times\text{propagators}.~~~~}
\end{align}
Here all dimensionful parameters are measured in the unit of $H$, and we have rewritten the $\de\phi$-correlation in terms of its diagrammatic representation. The vertices simply count the couplings in the unit of Hubble (and therefore, for example, a dimension-5 coupling suppressed by the cutoff $\Lambda$ is to be estimated as $H/\Lambda$). A loop factor will also be inserted if necessary. For propagators, the massless lines like inflaton fluctuations can be estimated as $\order{1}$.  To estimate massive propagators, we need to distinguish between hard momentum and soft momentum in the squeezed limit. A hard propagator with mass $m$ can be estimated by its flat-space value $1/m$ when $m\gg H$. On the contrary, when the propagator carries soft momentum and is responsible for generating the oscillatory signal, we should estimate it by the Boltzmann suppression factor $e^{-\pi m/H}$ when $m\gg H$.

We note that this estimate works only for the ``minimal scenario'' where all fluctuations are generated from the vacuum by expanding background. In the presence of other particle production mechanism (with a new scale $\mu>H$), such as warm inflation \cite{Tong:2018tqf} or resonant production \cite{Flauger:2016idt}, one should modify the estimates accordingly. We will consider such an example with chemical potentials in the next section. In the rest of this section, we will assume the minimal scenario without other production mechanisms.

\subsection{Estimate the signal sizes in simple inflation EFTs}

We now estimate the signal sizes in simple inflation scenarios with the method outlined above.
Instead of the details of the inflaton model, we focus on the size of the oscillating signal in the squeezed limit of the bispectrum due to the on-shell production of the new physics particles. In order for this to happen, new physics particles must be able to appear as intermediate states of the process with inflaton external lines. In addition, the physical mass of the new physics particle 
must be similar to the Hubble parameter. Otherwise, the signal would be suppressed exponentially $\sim \exp{(-\pi m/H)}$.  
Since the inflaton couples to matter field, its background value (both the field value and its time derivative) can shift of the mass spectrum of the matter field by an amount $\delta m$. The physical mass $m_{\rm{phys}}$ of a matter field is the sum of its ``original" mass and the shift induced by the inflaton. 
 For $m_{\rm{phys}} \sim H$ to happen naturally, we would like to have the induced mass shift to be at most $ H$.  As we will see, this turns out to be one of the most constraining requirements. 

\paragraph{Renormalizable couplings.}
We begin with the direct (namely non-derivative) couplings between the inflaton and matter fields. These couplings are renomalizable at leading order of EFT expansion. They break scale invariance, and so the coupling strength must be slow-roll suppressed. In the parameter regions we are interested in [See below in Eq.~(\ref{eq:tuning_renormalizable})], the back reactions to the inflaton potential from these couplings are always small.

 For a complex scalar matter field $Q$, the relevant Lagrangian is
\bge
\ld \supset -\mu \phi Q^\dagger Q -  \lam \phi^2 Q^\dagger Q - m_Q^2 Q^\dagger Q -  \lam_Q |Q|^4 - ({\mu}_Q Q^3 + \text{h.c.}).
\label{eq:renormalizable}
\ede
For a more general discussion, we have not assumed a $U(1)$ symmetry or a discrete symmetry associated with the matter field.\footnote{In addition to the $Q^3$ term, there can also be additional  couplings like $\lam_n Q^n+\text{c.c.}$~$(n=1,2,4,...)$ with complex $\lam_n$. Their presence won't affect the estimate we are doing. We ignore them for simplicity.  }  
In general,  $Q$ could acquire a VEV during the inflation, $\la Q \ra = Q_0$. The coupling in Eq.~(\ref{eq:renormalizable}) will generate time dependent contribution to the mass of $Q$ during the inflation. For example, term $\lam\phi^2Q^\dag Q$ introduces a time-dependent mass for $Q$, $\delta m^2_Q = \lambda \phi^2_0(t)$. This would be a new scale of non-adiabatic particle production, $\dot\omega\sim \dot m\sim \sqrt{\lam}\dot\phi_0$, where $\omega$ is the comoving energy of $Q$ being produced. To avoid Boltzmann suppression without fine tunings, we require various contributions to $Q$ mass being smaller than either $H^2$ or $\lam\dot\phi_0$,
\begin{eqnarray}
&\ &{\mathrm{Max}} \Big\{m_Q^2, \ \mu \phi_0, \  \lambda \phi_0^2, \ \lam_Q Q_0^2, \ \mu_Q Q_0 \Big\} \lesssim \text{Max}\Big\{H^2,\lam\dot\phi_0\Big\}.
\end{eqnarray}
In principle, the condition depends on the relative sizes of $H^2$ and $\lam\dot\phi_0$. However, we note that $\lam\dot\phi_0>H^2$ is quite special as it introduces a strong scale dependence. In this case a large signal could be produced when $\sqrt{\lam}\phi^2_0<\dot\phi_0$. 
Even though we expect $\phi_0$ to be much larger than $\sqrt{\dot\phi_0}$ in a generic point of the inflation trajectory (see below), there are situations when this condition is fulfilled. In this special case there can be scale-dependent features in the power spectrum as well as in 3-point correlations. There can be associated oscillating signals in the squeezed bispectrum, too. See \cite{Flauger:2016idt} for an example. In the following we will only consider the alternative case with $\lam\dot\phi_0<H^2$ so that the scale invariance is not strongly affected. In this case we will have the following conditions on various parameters,
\begin{align}
 \mu < \frac{H^2}{\phi_0}, \ \lambda < \left( \frac{H}{\phi_0} \right)^2, \ \lambda_Q < \left( \frac{H}{Q_0}\right)^2 , \ \mu_Q < \frac{H^2}{Q_0}. 
 \label{eq:tuning_renormalizable}
\end{align}

We first observe that there is no tree-level contribution to the signal. There is a tree-level contribution to the NG. However, in order to have the oscillatory signal, we need to be able to have the new physics particle $Q$ as on-shell intermediate state. For this to be the case for a tree-level 3-point function, we need a non-zero 2-point interaction between the inflaton and $Q$. For the renormalizable couplings with no derivatives, as those in Eq.~(\ref{eq:renormalizable}), the only possible two point functions are mass mixing and kinetic mixing between the inflaton and $Q$, both of which can be diagonalized and does not lead to physical 2-point inflaton-$Q$ coupling.\footnote{One can also see this explicitly in the diagrammatic level by treating the mass mixing perturbatively: Let $\si_{1,2}$ be two massive real scalars with masses $m_{1,2}$ in inflation. Then the 2-point mass mixing diagram connecting the propagators $G_{1,2}$ of both fields can be decomposed as $\int \di^4y\sqrt{-g(y)}G_1(x,y)G_2(y,z)=-[G_1(x,z)-G_2(x,z)]/(m_1^2-m_2^2)$. See Appendix A.4 of \cite{Chen:2016hrz} for more details. } Hence, there can only be contributions to the signal at 1-loop order. For example, one could have
\bge
f_\text{NL}^\text{(osc)}\left(\parbox{0.3\textwidth}{\includegraphics[width=0.3\textwidth]{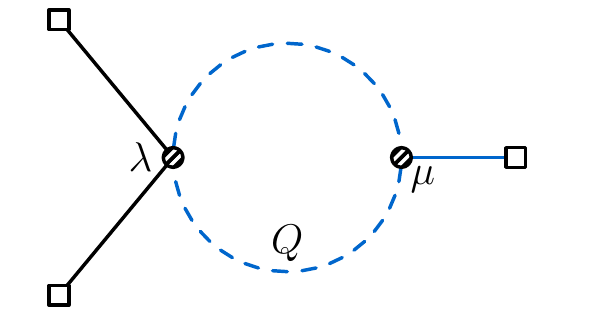}}\right) \sim \FR{1}{2\pi P_\zeta^{1/2}}\cdot \frac{1}{16 \pi^2} \lam \frac{\mu}{H} < \FR{1}{2\pi P_\zeta^{1/2}} \frac{1}{16 \pi^2} \left( \frac{H}{\phi_0}\right)^3.
\label{eq:NG1}
\ede
We have adopted the diagrammatic notations introduced in \cite{Chen:2017ryl}, where squares denote external (boundary) points and shaded circles denote bulk vertices. Here and below we use blue lines to denote propagators carrying the soft momentum.

To proceed further, we need an estimate of the typical field value $\phi_0$ of the inflaton. We note that there needs to be about 10 $e$-folds just for the CMB,
\bge
N \simeq \int \di \phi\,\FR{H}{\dot\phi_0} \sim   \FR{H}{\dot\phi_0} \Delta\phi = {\cal O} (10),
\ede
where we have assumed that slow-roll parameter $\epsilon$ remains approximately constant during the  epoch when the fluctuations relevant for the CMB were generated. Generically, we also expect\footnote{It is of course possible that $\phi_0 \ll \Delta \phi$ at some special points in the field trajectory of the inflaton. This could allow us to have a smaller contribution to the mass of the matter particle.  However, the argument here shows that it would be difficult maintain it during the full period of inflation relevant for CMB. }   $\Delta \phi \sim \phi_0$.   Therefore, we have 
\bge
\phi_0 \sim \Delta \phi \sim N \dot\phi_0 /H \ \To \ \frac{H}{\phi_0} \sim  \frac{2 \pi P_\zeta^{1/2}}{ N} \sim 10^{-5}. 
\label{eq:fieldrange}
\ede
Hence, from Eq.~(\ref{eq:NG1}), we have 
\bge
f_\text{NL}^\text{(osc)} \sim  \frac{1}{4N^3} P_\zeta 
\ede
which is too small to be observed. Other contributions, such as the ones present after $Q$ acquiring a VEV,  can be estimated similarly. However, they do not change the conclusion qualitatively. 

One could also consider renormalizable coupling between the inflaton and new physics fermions or gauge bosons
\bge
\ld \supset y \phi \bar \Psi \Psi + |D_\mu \Phi|^2.
\ede
For the coupling to fermions, requiring $m_{\Psi} \sim H$ and lack of fine-tuning leads to $y < H/\phi_0$. For the coupling to gauge bosons, the inflaton needs to be part of a complex scalar which is charged under the gauge interaction. The gauge boson will have mass $m_A^2 \sim g^2 \phi^2_0$. Then, we will need to have $g \lesssim H/\phi_0$ to have a gauge boson which is naturally around the Hubble scale. In both of these cases, the contribution to the bispectrum is only at one-loop level. An estimate similar to that of the scalar case gives similar conclusions.

\paragraph{Derivative couplings.}
Next, we move on to consider non-renormalizable couplings. Here, with the inflation background, it is possible to have  two point couplings of the form $\dot{\de\phi} Q$,  which are necessary to have tree-level contribution to the oscillatory signal.    A distinct possibility is that the inflaton only couples derivatively, which would be the case if it has an approximate shift symmetry $\phi \to \phi + c$. So the signals produced from these couplings will be nearly scale invariant up to slow-roll corrections.

A commonly considered example in this category is that the inflaton couples to matter field through a dimension-6 operator. The relevant Lagrangian is
\bge
\ld \supset \frac{c_6}{\Lambda^2} (\pd \phi)^2 Q^\dagger Q - m_Q^2 Q^\dagger Q - \lam_Q |Q|^4,
\label{eq:o6}
\ede
where $\Lambda$ is the scale of the sector which mediates the interaction. If the mass of $Q$ is comparable to $H$, this is the so called quasi-single field inflation \cite{Chen:2009zp,Baumann:2011nk,Kumar:2017ecc}. The resulting NG is larger if the matter field acquires a VEV, $\la Q \ra = Q_0$. During the inflation, $\la \pd_\mu \phi \ra = \phidot\de_{\mu 0} $, there is a correction to the mass of $Q$. To avoid fine-tuning
\bge
{\rm{Max}}\bigg\{m_Q^2, \ \lam_Q Q_0^2, \ \delta m^2_Q = c_6 \Big(\frac{\phidot}{\Lambda}\Big)^2 \bigg\}  \lesssim H^2,
\ede
which means
\bge
\frac{c_6}{\Lambda^2} \lesssim \frac{H^2}{\phidot^2} = \FR{(2 \pi)^2P_\zeta}{H^2}.
\ede
There are three tree-level contributions to NG:
\bge
 \parbox{0.23\textwidth}{\includegraphics[width=0.23\textwidth]{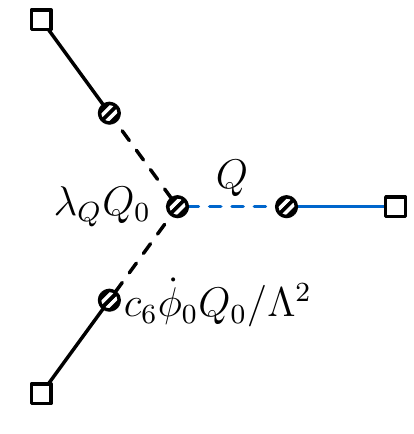}}
 \parbox{0.23\textwidth}{\includegraphics[width=0.23\textwidth]{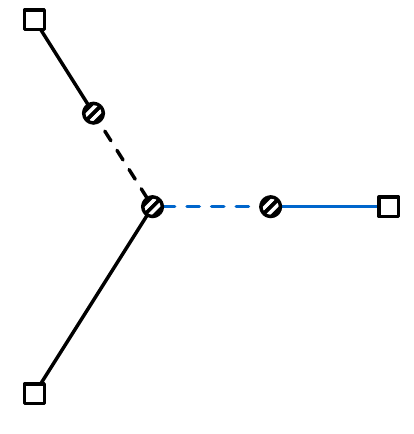}}
 \parbox{0.23\textwidth}{\includegraphics[width=0.23\textwidth]{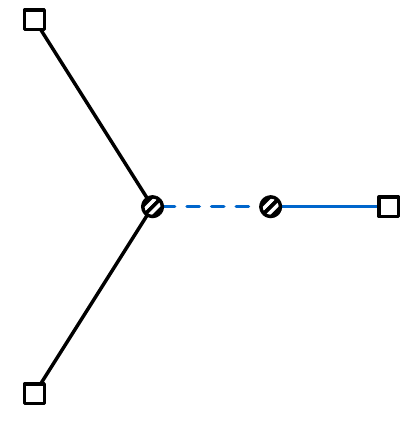}}
\ede
In the most favorable case which involves cubic self-coupling of $Q$,  
\begin{align}
\fo \left(\parbox{0.18\textwidth}{\includegraphics[width=0.18\textwidth]{Fig_3ptQSFI1}}\right)\sim \FR{1}{2\pi P_\zeta^{1/2}} \Big( \frac{c_6}{\Lambda^2} \phidot  Q_0 \Big)^3 \lam_Q Q_0  \frac{1}{H^4}
 \lesssim \lam_Q^{-1}(2\pi)^2P_\zeta .
\end{align}
Hence, to have a observable signal $f_\text{NG}\sim\order{1}$, a small quartic coupling $\lam_Q \sim (2\pi)^2P_\zeta\simeq 8\times 10^{-8}$ (or equivalently, a large VEV $Q_0 \sim 3\times 10^3 H$\footnote{We note that such a VEV of $Q_0$ is already larger than the minimal value of the cut-off $\Lambda \sim \sqrt{\dot \phi_0}$. Hence, the assumption of being able to ignore high order terms in the EFT expansion implies additional fine tuning in this scenario. }. ) is necessary. We note that the coefficient of the two-point mixing $c_6\dot\phi_0Q_0/\Lambda^2$ is comparable to $H$ in this regime and thus $\de\phi$ and the massive field $Q$ are quite strongly coupled. This is not a problem since the strongly coupled regime of quasi-single field inflation is well understood \cite{An:2017hlx,Iyer:2017qzw}, and the above estimate work reasonably well when the mixing is of $\order{H}$.

 There is a dim-7 operator of similar form with inflaton coupling to fermion bilinear. However, this only contributes to NG at one-loop order which is smaller.

Finally, we consider dim-5 couplings of the form 
\bge
\frac{1}{\Lambda} \pd_\mu \phi \mathcal{J}^\mu, \ \frac{\phi}{\Lambda} F \wedge F.
\label{eq:o5}
\ede
For the first operator, $\mathcal{J}^\mu$ is the current associated with a symmetry for which is non-linearly realized by the shift of $\phi$. The second operator is a typical coupling between an axion-like-particle and gauge (gravity) field strengths.  A detailed understanding of the effect of this class of operators is a main topic of this paper. Here, to set the stage, we first present a simple (and naive) estimate of its contribution to NG. We note that an important feature of this class of operators is that they do not directly contribute to the mass spectrum of the matter fields, including the constituents of $\mathcal{J}_\mu$ and the gauge field. There is still a constraint on the cut-off scale from the validity of the EFT expansion
\bge
\phidot < \Lambda^2. 
\ede
This class of operators do not contribute to the bispectrum at the tree level. Naively, their contribution can be estimated as 
\bge
\fo \sim \frac{1}{16 \pi^2}  \FR{1}{2\pi P_\zeta^{1/2}}\Big( \frac{H}{\Lambda} \Big)^3 < \frac{1}{16 \pi^2} \FR{1}{2\pi P_\zeta^{1/2}}\bigg( \frac{H}{\phidot^{1/2}} \bigg)^3 = \FR{1}{16\pi^2}\cdot \sqrt{2\pi} P_\zeta^{1/4}.
\ede
This has a smaller power in $P_\zeta$ even though one has to pay an additional loop factor. In comparison with Eq. (77) of Ref.~\cite{Chen:2018xck}, this agrees parametrically with the pre-factor $\phidot^{1/2}$ if we saturate the EFT limit $\Lambda = \phidot^{1/2}$.  The final result, Eq. (79) of of Ref.~\cite{Chen:2018xck}, features additional enhancement. Understanding them will be the focus of next sections.

\paragraph{Trispectrum.} To finish this section we also briefly consider the trispectrum. The estimate of the 4-point non-Gaussian parameter $T$ goes as 
\begin{align}
  T\sim \FR{1}{(2\pi)^2P_\zeta}\times \text{loop factor}\times\text{vertices}\times \text{propagators}.
\end{align}
The trispectrum is more challenging to probe in general, but there are cases where the signals show up only in the trispectrum but not in the bispectrum. For example, we have shown that the renormalizable couplings can not show up at the tree level because of the lack of a nontrivial two-point mixing. Nevertheless, renormalizable couplings can give tree-level trispectrum, although the amplitude is again tiny. Take $\lam \phi^2 Q^\dagger Q$ as an example, we have
\bge
T \sim \frac{1}{(2\pi)^2P_\zeta} \lam^2 \frac{Q_0^2}{H^2} <  \frac{1}{(2\pi)^2P_\zeta} \bigg( \frac{H}{\phi_0}\bigg)^4 \frac{Q_0^2}{H^2}  = \frac{(2 \pi)^2P_\zeta}{N^2}   \frac{Q_0^2}{H^2}
\ede
where we have used the no-fine-tuning condition in Eq.~(\ref{eq:tuning_renormalizable}) and the condition on the field range in Eq.~(\ref{eq:fieldrange}). The signal is very small unless $Q_0 \gg H$. 

Another possibility would be to have a coupling of the form $g A_\mu \pd^\mu Q \phi$. This would be the case, for example, if inflaton $\phi$ is part of a complex scalar $\Phi$. At the same time,  $\Phi$ and $Q$ both charged under the gauge interaction mediated by $A_\mu$. In this case, we still expect $m_A \sim g \phi_0$, hence $g < H/\phi_0$ if we would like to have $A$ naturally lighter than $H$. We can further assume there is a mixing between $Q$ and $\phi$ of the form $\pd \phi Q$ much like the case of dimension 6 operator in Eq.~(\ref{eq:o6}), $\sim v {\phidot}/{\Lambda^2}$. In this case, there is a tree-level contribution to the trispectrum with the gauge boson as an intermediate state. We can estimate
\bge
T \sim \frac{1}{P_\zeta} g^2 \left(v \frac{\phidot}{\Lambda^2} \right)^2 \frac{1}{H^2} < \frac{(2 \pi)^4}{N^2} P_\zeta.
\ede

\section{Intuitive Understanding of Chemical Potential}
\label{sec_intuitive}

During inflation, a nonzero chemical potential is introduced by the rolling inflaton field via the dim-5 inflaton-matter couplings shown in (\ref{eq:o5}). It was previously known that such a chemical potential may lead to enhanced particle production when the matter fields are inflated outside the horizon. It was also known that not all such chemical potentials do the job. Here we would like to understand the underlying physical reason.  

The basic physical picture can already be seen in the flat space limit. So we will temporarily ignore the cosmic expansion. Generally speaking, we can introduce a chemical potential whenever we have an additive charge density of $\mathcal{Q}$. A nonzero chemical potential $\mu$ is introduced by shifting the Hamiltonian of the system as $\mathcal{H}\to \mathcal{H}-\mu \mathcal{Q}$. Then, a positive $\mu$ will tend to produce more positively charged particles in order to enhance $\mathcal{Q}$ and thus to minimize the free energy. 

Consider a simple example of a free massive complex scalar field $\Phi$ charged under a global $U(1)$. The corresponding conserved charge is $\mathcal{Q}=-\ii(\dot\Phi^*\Phi-\Phi^*\dot\Phi)$. From this we can find the Lagrangian $\ld$ by finishing the momentum integral in the partition function $\mathcal{Z}=\int\Di\Phi\Di\Pi (- \int \di^4x(\mathcal{H}-\mu\mathcal{Q}))=\int\Di\Phi(- \int\di^4x\ld)$. The result is
\begin{align}
\label{scalarLag}
  \ld=[(\pd_t+\ii\mu)\Phi^*][(\pd_t-\ii\mu)\Phi]-|\pd_i\Phi|^2-m^2|\Phi|^2.
\end{align}
However the chemical potential $\mu$ here has no (independent) observable effects in inflation. To understand it, we note that the chemical potential $\mu$ is eliminated by a field redefinition $\Phi\to e^{\ii\mu t}\Phi$. When we decompose $\Phi$ into modes $\Phi\sim\int\di^3k e^{-\ii\omega t+\ii\mb k\cdot\mb x}(a_{\mb k}+b_{\mb k}^\dag)$,
the field redefinition amounts to a replacement $\omega\to \omega-\mu$. (Here we are thinking of a negative mode as a mode with negative energy $-\omega$.) So the field redefinition is nothing but reclassifying positive and negative modes, and this reclassification is independent of time. So, if we begin with a Bunch-Davis vacuum where only the (reclassified) positive modes are excited, the mode will evolve as if there is no chemical potential at all. 

The second way to understand the absence of a physical effect is to note that the chemical potential in (\ref{scalarLag}) is formally equivalent to a nonzero temporal component of a gauge potential $A_\mu$. Then the field redefinition amounts to a constant shift in $A_0$. The zero point of the electric potential $A_0$ can of course be chosen arbitrarily and this choice should have no physical consequences. 

A similar case is a Dirac fermion $\Psi$ with vectorial $U(1)$ symmetry. Here the conserved current is $\mathcal{J}^\mu=\ob\Psi\ga^\mu\Psi$ and thus the conserved charge is $\mathcal{Q}=\ob\Psi\ga^0\Psi$. Inserting this into the Lagrangian,
\bge
  \ld = \ob\Psi(\ii\ga^\mu\pd_\mu-\mu\ga^0-m)\Psi,
\ede
we see again that the chemical potential term is eliminated by a field redefinition $\Psi\to e^{-\ii\mu t}\Psi$. The physical interpretation is identical to the previous one.%
\footnote{This line of arguments seem to suggest that the chemical potentials in these examples are completely redundant since the path integral can always be made independent of $\mu$. This is somewhat confusing since we know a nonzero chemical potential in a thermal equilibrium system does have a nontrivial effect, such as Bose-Einstein condensation for scalar field. This confusion is removed if we note that the partition function in thermal equilibrium is related to the above path integral by Wick rotation. Then the resulting kinetic term will be like $[(\pd_\tau+\mu)\Phi^*][(\pd_\tau-\mu)\Phi]$. The chemical potential $\mu$ here can no longer be removed by a simple field redefinition. Another way to see this is that the chemical potential still introduces a constant shift in Matsubara frequency but it is pure imaginary $\omega\to \omega+\ii\mu$ so can in no way be removed by a redefinition of the zero point. }

In both cases above, the effect of the chemical potential can be understood as a constant shift in the frequency $\omega$ of the mode function, so that the dispersion relation of the matter field is
\bge
\label{disp1}
(\omega\pm\mu)^2=k^2+m^2.
\ede
Therefore it is a redefinition of positive frequency and negative frequency. In inflation, the chemical potential $\mu$ is usually a constant and thus this redefinition is independent of time. So no physical effects will appear.

At this point, it is helpful to recall what do we mean by a physical effect: it is the enhanced particle production during inflation. Technically, the particle production is possible when WKB approximation fails to capture the evolution of modes. The modification of the dispersion relation in (\ref{disp1}) does not change the evolution of the mode, so we get no effects in particle production. This motivates us to look at an alternative modification of dispersion, namely a constant shift in the  \emph{physical momentum},
\bge
\label{disp2}
 \omega^2=(k\pm \mu)^2+m^2+\cdots,
\ede
where the dots represent possible constant correction from the chemical potential. This type of chemical potentials can usually generate physical effects in inflation. The reason is clear: the physical momentum $k$ during inflation is experiencing exponentially fast redshift, $k\sim a^{-1}\sim e^{-Ht}$. For a mode with constant \emph{comoving} momentum $k_\text{com}$, the effect of $\mu$ looks like a time dependent mass,
\bge
 \omega^2=a^{-2}k_\text{com}^2+\Big[m^2\pm 2\mu k_\text{com}a^{-1}(t)+\cdots\Big],
\ede
where the dots represent time-independent corrections. This time-dependent mass will change the evolution of the mode, and also the efficiency of particle production. Therefore, we should look for chemical potentials that modify the dispersion relation according to (\ref{disp2}).

To see which type of chemical potential can lead to (\ref{disp2}), we note that a constant shift in $k$ generates a term linear in $k$. In the context of EFT, this linear term can appear only from dotting the 3-momentum $\mb k$ with some other 3-vectors $\mb n$. So we know immediately that no such chemical potential is available for scalars without breaking the 3-rotation symmetry, due to the lack of a preferred direction $\mb n$. We note in passing that the particle production with broken 3-rotation has been studied in \cite{Chua:2018dqh}.

To realize (\ref{disp2}) without breaking 3-rotation, we need to look at fields with nonzero spin, since the momentum and the spin are the only (Casimir) invariant vectors for a particle. In this case, we have a spin vector $\mb s$ from the field itself and we can use it to form a scalar linear in $\mb k$, namely $\mb k\cdot \mb s$. This quantity has odd parity. To realize it from a chemical potential, we will need a pseudo-vector conserved current. With spin-1/2 and spin-1 fields, two such examples are chiral current $\ob\Psi\ga^5\ga^\mu\Psi$ and the Chern-Simons current $\ep^{\mu\nu\rh\si}A_\nu\pd_\rh A_\si$. It is already known that both of them lead to enhanced particle production in inflation and  potentially large cosmological collider signals. We will study them more carefully in next section. 

In summary, we have shown in this section that the chemical potential from dim-5 couplings could enhance particle production during inflation only when it is associated with a parity-odd particle number density, if we do not break scale invariance and 3-rotation. In next section we will justify this conclusion by calculating the mode functions of heavy particle explicitly. By doing so we will also have a more quantitative understanding of the signal strength from the chemical potential.

\section{Mode Functions with Chemical Potentials}
\label{sec_mode}

In this section, we supplement the arguments in previous sections by calculating explicitly the mode functions in several examples in inflation.

During inflation the inflaton $\phi$ is slowly rolling. In standard FRW coordinates, the $\pd_\mu\phi$ acquires a VEV $\la\pd_\mu\phi\ra=\dot\phi_0 \de_{0\mu}$. We can couple $\pd_\mu\phi$ to a current $\mathcal{J}^\mu$. Then the operator $(\pd_\mu\phi)\mathcal{J}^\mu/\Lambda$ becomes $\mu\mathcal{J}^0$ where $\mu=\dot\phi_0/\Lambda$ is effectively a chemical potential associated with the ``charge density'' $\mathcal{J}^0$. As discussed in Sec.~\ref{sec_size}, the validity of the EFT expansion requires roughly that $\Lambda>\dot\phi_0^{1/2}$. Therefore, we have  $\mu < \dot\phi_0^{1/2}$. Given $\dot\phi_0\simeq 60H$ in ordinary inflation scenarios, we see that the chemical potential $\mu$ is large a priori compared with the Hubble. 

In the context of EFT, the operators generating chemical potentials are among the leading order ones of a dimensional expansion of EFT operators when imposing a shift symmetry $\phi\to \phi+$const. to the inflaton. In the following, we consider cases where the current $\mathcal{J}^\mu$ is made of scalars, fermions, gauge bosons, and finally graivtons, respectively.  

\subsection{Scalar}
 A single real scalar boson cannot have  a finite chemical potential since  it can not have a continuous $U(1)$ symmetry.  We need at least two real scalars (or a complex scalar), in addition to the inflaton. 
 Let $\si_{1,2}$ be two real scalars. In general, they can have different masses $m_{1,2}$, respectively. The lowest dimensional couplings of $\si_{1,2}$  to the inflaton are from the operators, 
\begin{align}
  \Delta\ld=\sqrt{-g}\bigg[\FR{c_1}{\Lambda}(\pd_\mu\phi)(\pd^\mu\si_1)\si_2+\FR{c_2}{\Lambda}(\pd_\mu\phi)(\pd^\mu\si_2)\si_1\bigg].
\end{align}
During inflation, the above coupling becomes
\begin{align}
  \Delta\ld=-a^3\big(\mu_1\si_1'\si_2+\mu_2\si_2'\si_1\big),
\end{align}
with $\mu_{1,2}\equiv c_{1,2}\dot\phi_0/\Lambda$.  It is also convenient to define combinations $\si_\pm=\si_1\pm \ii\si_2$.  In the limit of $m_1 = m_2$ and $\mu_1= -\mu_2$, there is an unbroken $U(1)$ symmetry under which $\si_\pm \to e^{\pm i \alpha }\si_\pm$. With a local $U(1)$ rotation, we can remove the dimension-5 inflaton couplings. Hence, we will investigate the most general case in which $m_1 \neq m_2$ and $\mu_1 \neq - \mu_2$. 

The equations of motion for the $\mb k$ modes of $\si_\pm$ are
\begin{align}
\si_\pm''-\FR{2}{\tau}\bigg(1\pm\FR{\ii\mu_-}{H}\bigg)\si_\pm'+\bigg(k^2+\FR{m_+^2}{H^2\tau^2}\pm\FR{3\ii\mu_-}{H\tau^2}\bigg)\si_\pm + \bigg(\FR{m_-^2}{H^2\tau^2}\mp\FR{3\ii \mu_+}{H\tau^2}\bigg)\si_\mp=0 .
\end{align}
where $m_\pm^2\equiv(m_1^2\pm m_2^2)/2$ and $\mu_\pm\equiv(\mu_1\pm \mu_2)/2$. 
To proceed, we first define $\si_\pm=f_\pm\chi_\pm$ to eliminate the first-order derivative terms. Choosing $f_\pm=\tau^{1\pm\ii\mu_-/H}$, we get,
\begin{align}
  \chi_\pm''+\bigg(k^2+\FR{m_+^2+\mu_-^2-2H^2}{H^2\tau^2}\bigg)\chi_\pm+\FR{m_-^2\mp3\ii H \mu_+}{H^2\tau^2}\tau^{\mp\ii 2 \mu_- /H}\chi_\mp=0.
\end{align}
Then we can use a unitary rotation of $\chi_\pm$ to diagonalize the non-derivative terms. Denoting the diagonal basis by $\wt\chi_{\pm}$, we have
\begin{align}
  \wt\chi_{\pm}''+\Bigg(k^2+\FR{m_+^2+\mu_-^2-2H^2\pm\sqrt{m_-^4+9H^2\mu_+^2}}{H^2\tau^2}\Bigg)\wt\chi_{\pm}=0.
\end{align}
The solution is
\begin{align}
\label{sigmamode}
  \wt{\chi}_{\pm}(\mb k,\tau)=-\FR{\ii\sqrt{\pi}}{2}e^{\ii\pi(\nu_\pm/2+1/4)} H(-\tau)^{1/2}\text{H}_{\nu_\pm}^{(1)}(-k\tau),
\end{align}
with parameter $\nu_\pm=\sqrt{9/4-(\wt m_\pm/H)^2}$ and $\text{H}_{\nu}^{(1)}$ is H\"ankel function of first kind. The effective masses $\wt m_\pm$ of the two modes $\wt\chi_\pm$ are
\begin{align}
\label{masseigenvalue}
  \wt m_\pm^2=m_+^2+\mu_-^2\pm\sqrt{m_-^4+9H^2\mu_+^2}.
\end{align}
From this, we can see the dependence on the chemical potential in the mode function $\si_\pm$ are through parameter $\nu_\pm$ and factor $(-\tau)^{3/2 \pm \ii \mu_-/H}$. There is no enhancement which can potentially compensate the suppression $e^{-\pi m/H}$. 

One special case is the possibility of $\wt m_-^2<0$. In this case, $\wt\chi_-$ will be a growing mode during inflation due to the tachyonic instability. This growing mode will not lead to an enhanced clock signal, but there could be interesting phenomenological consequences which we will study in a future work.

\subsection{Spin-1/2 fermion} 

We have argued that the chemical potential will only have a physical effect while associated with a chiral current. Hence, we begin by considering a 2 component Weyl fermion $\psi$.
The chemical potential for $\psi$ can be introduced from the dim-5 operator $(\pd_\mu\phi)\psi^\dag\ii\bar\si^\mu\psi/\Lambda$. This case has been studied in \cite{Chen:2018xck}. See also \cite{Hook:2019zxa}. Here we will summarize the main result following \cite{Chen:2018xck} with slightly different notations. We first evaluate the Lagrangian
\begin{align}
  \ld=\sqrt{-g}\bigg[\ii\psi^\dag\ob\si^\mu\D_\mu\psi-\FR{1}{2}m(\psi\psi+ \psi^\dag \psi^\dag)+\FR{1}{\Lambda}(\pd_\mu\phi)\psi^\dag \ob\si^\mu\psi\bigg],
\end{align}
with the inflaton background, which then becomes
\begin{align}
\label{fermionld}
  \ld=\ii\psi^\dag\ob\si^\mu\pd_\mu\psi-\FR{1}{2}am(\psi\psi+\psi^\dag\psi^\dag)-a\mu\psi^\dag\ob\si^0\psi.
\end{align}
Here we have defined the chemical potential $\mu=\dot\phi_0/\Lambda$ and made the field redefinition $\psi\to a^{-3/2}\psi$ which turns the covariant derivative $\D_\mu\psi$ into partial derivative $\pd_\mu\psi$. Then we decompose $\psi$ into modes,
\bge
\label{mode}
   \psi_\al(\tau,\mb x)= \int\FR{\di^3\mb k}{(2\pi)^3}\sum_{s=\pm}\Big[\xi_{\al,s}(\tau,\mb k)b_s(\mb k)e^{+\ii\mb k\cdot\mb x}+\chi_{\al,s}(\tau,\mb k)b_s^{ \dag}(\mb k)e^{-\ii\mb k\cdot\mb x}\Big],
\ede
where we have spelled out the spinor indices explicitly. The mode function $\xi$ and $\chi$ can be further written in terms of normalized helicity eigenstates $h_s$, defined by $\vec\si\cdot\vec k h_s(\mb k)=s k h_s(k)$ and the usual orthonormal condition, as
\begin{align}
\label{helimode}
  &\xi_{\al,s}(\tau,\mb k)=u_s(\tau,\mb k)h_s(\mb k), && \chi^{\dag\dot \al}_s(\tau,\mb k) = v_s(\tau,\mb k) h_s(\mb k).
\end{align}
Then the equation of motion for the modes $u$ and $v$ can be derived as
\bgs
\label{fermioneom}
\begin{align}
  &\ii u_\pm' \pm k u_\pm = a\mu u_\pm + am v_\pm ,\\
  &\ii v_\pm'\mp k v_\pm =-a\mu v_\pm +am u_\pm.
\end{align}
\eds
These two equations can be decoupled into a pair of second order equations,
\begin{align}
  &u_\pm''-a H u_\pm'+\Big[(k\mp a\mu)^2+a^2m^2\pm\ii aH k\Big]u_\pm = 0,\\
  &v_\pm ''-a H v_\pm'+\Big[(k\mp a\mu)^2+a^2m^2\mp\ii aH k\Big]v_\pm = 0.
\end{align}
One can see directly the flat-space dispersion relation from this equation by putting $a=1$ and $H=0$, 
\bge
  \omega^2=(k\pm \mu)^2+m^2.
\ede
During inflation, the wavenumber $k$ here should be taken as the physical wavenumber which is $k_\text{phys}=k_\text{com}/a$ and is being redshifted as $a$ grows exponentially. The total energy $\omega$ is minimized in this evolution when $k_\text{phys}=\mu$, at which point the adiabatic approximation fails maximally. This gives a new scale of the particle production, above the usual one due to cosmic expansion that happens at $k_\text{phys}\simeq H$. On the other hand, we note that the energy $\omega$ here is always positive for all values of $\mu$. This is not unrelated to the Fermi statistics, in which case the Pauli blocking forbids any possible instability that could be potentially introduced by the chemical potential.

The effect of the chemical potential goes away for massless fermions.  In this case, the Lagrangian in (\ref{fermionld}) has a chiral symmetry. We can use a local  chiral rotation,  $\psi\to e^{\ii\phi/\Lambda}\psi$  to eliminate the chemical potential. On the other hand, if $m \neq 0$, the same chiral rotation will reintroduce the chemical potential as a complex phase of the mass term, $m\to m e^{2\ii\phi/\Lambda}$. 

The effect of the chemical potential can be seen explicitly in the mode function of the fermion. 
The equations (\ref{fermioneom}) can be solved directly with the usual initial condition that only positive frequency modes are excited. The solutions are 
\bge
\begin{aligned} 
  &u_{+}(\tau,\mb k)= \FR{(m/H) e^{+\pi\mu/(2H)}}{\sqrt{-2k\tau}}\mathrm{W}_{\ka,\ii \nuf}(2\ii k\tau) , 
  &&u_{-}(\tau,\mb k)= \FR{e^{-\pi\mu/(2H)}}{\sqrt{-2k\tau}}\mathrm{W}_{-\ka,\ii \nuf}(2\ii k\tau) ,\\
  &v_+(\tau,\mb k)= \FR{e^{+\pi\mu/(2H)}}{\sqrt{-2k\tau}}\mathrm{W}_{1+\ka,\ii \nuf}(2\ii k\tau) , 
  &&v_-(\tau,\mb k)= \FR{(m/H) e^{-\pi\mu/(2H)}}{\sqrt{-2k\tau}}\mathrm{W}_{-1-\ka,\ii \nuf}(2\ii k\tau),
\end{aligned}
\ede
where $\mathrm{W}_{a,b}(z)$ is the Whittaker function, with $\ka=-1/2-\ii\mu/H$ and $\nuf=\sqrt{m^2+\mu^2}/H$. 

For cosmological collider signals we need the late time behavior of these mode functions, 
\label{mfltl}
\begin{align}
 u_+(\tau,\mb k)\simeq &~ e^{-\ii\pi/4}e^{+\pi\mu/(2H)}\FR{m}{H}\bigg[\FR{e^{\pi\nuf/2}\Gamma(-2\ii\nuf)}{\Gamma(1+\ii\mu/H-\ii\nuf)}(-2k\tau)^{\ii\nuf}+(\nuf\to-\nuf)\bigg],\\
 u_-(\tau,\mb k)\simeq &~ e^{-\ii\pi/4}e^{-\pi\mu/(2H)} \bigg[\FR{e^{\pi\nuf/2}\Gamma(-2\ii\nuf)}{\Gamma(-\ii\mu/H-\ii\nuf)}(-2k\tau)^{\ii\nuf}+(\nuf\to-\nuf)\bigg],\\
 v_+(\tau,\mb k)\simeq &~ e^{-\ii\pi/4}e^{+\pi\mu/(2H)} \bigg[\FR{e^{\pi\nuf/2}\Gamma(-2\ii\nuf)}{\Gamma(\ii\mu/H-\ii\nuf)}(-2k\tau)^{\ii\nuf}+(\nuf\to-\nuf)\bigg],\\
 v_-(\tau,\mb k)\simeq &~ e^{-\ii\pi/4}e^{-\pi\mu/(2H)}\FR{m}{H}\bigg[\FR{e^{\pi\nuf/2}\Gamma(-2\ii\nuf)}{\Gamma(1-\ii\mu/H-\ii\nuf)}(-2k\tau)^{\ii\nuf}+(\nuf\to-\nuf)\bigg].
\end{align}
These are good approximations when $|k\tau|<1$. On the other hand, the oscillatory bispectrum receives most of its contribution from the region $|k\tau|\sim \nuf$. When $\nuf\gg 1$ which is phenomenologically relevant, an improved approximation is needed which could introduce an $\order{1}$ difference \cite{Hook:2019zxa}. We will not go into the details in this study. 

The region $\mu\gg m\gg H$ is particularly interesting because some components of the mode functions are not exponentially suppressed. One can understand this by noting that the chemical potential $\mu$ leads to exponentially enhanced particle production on the one hand, and also an effective contribution to the mass $\nuf^2H^2=m^2+\mu^2$ which leads to Boltzmann suppression on the other hand. Combining the two exponential factors, we have $e^{+\pi\mu/H}e^{-\pi\nuf}\simeq e^{-\pi m^2/(2\mu H)}$. So the mode function can be taken very roughly as $\order{1}$ when estimating cosmological collider signals if $\mu\gg m\gg H$. More details are in the next section.

The generalization of the discussion above to a Dirac fermion, $\Psi=(\chi,\xi^\dag)^T$,  is straightforward. A generic fermion current can be written as $\mathcal{J}^\mu = g_V \mathcal{J}_V^\mu + g_A \mathcal{J}^\mu_A$, where $\mathcal{J}_V = \Psi\ga^\mu\Psi $ and $\mathcal{J}_A = \Psi \gamma^5 \ga^\mu\Psi $. As we have already discussed in Sec.~\ref{sec_intuitive}, a coupling $\pd_\mu \phi \mathcal{J}_V^\mu$ can be removed via a local field redefinition.  Hence, we only need to consider the coupling to the axial current $\pd_\mu \phi \mathcal{J}_A^\mu$. In terms of Weyl spinors, it can be rewritten as
\begin{align}
  \FR{1}{\Lambda}(\pd_\mu\phi)\ob\Psi\ga^5\ga^\mu\Psi=\FR{1}{\Lambda}(\pd_\mu\phi)(\chi^\dag\ob\si^\mu\chi+\xi^\dag\ob\si^\mu\xi)=\FR{1}{\Lambda}(\pd_\mu\phi)(\psi_1^\dag\ob\si^\mu\psi_1+\psi_2^\dag\ob\si^\mu\psi_2).
\end{align}
This reduces to two decoupled copies of the Lagrangian (\ref{fermionld}), and similar conclusions apply.

\subsection{Gauge boson}

The lowest dimensional inflaton-gauge boson coupling respecting the shift symmetry is the dim-5 operator $-\phi F\wt F/(4\Lambda)$. The treatment below follows \cite{Lu:2019tjj}. Let the gauge boson have mass $m$. The equation of motion in the covariant gauge $\pd_\mu(\sqrt{-g}A^\mu)=0$ is
\begin{equation}
    \mb A''+k^2\mb A+a^2m^2\mb A-\ii a\mu\mb k\times \mb A=0,
\end{equation}
where the chemical potential $\mu=\dot\phi_0/\Lambda$. Separating the gauge field into transverse and longitudinal polarizations $(A_\pm,A_3)$ with $A_\pm = \frac{1}{\sqrt 2} (A_1\pm \ii A_2)$, we get the equations for the modes,
\begin{align}
\label{Aeom}
    & A_\pm''+(k^2 +a^2m^2 \pm 2a\mu k )A_\pm=0, 
    &&A_3''+(k^2+a^2m^2)A_3=0.
\end{align}
We can now go again into the physical time and find the dispersion relation of the gauge boson, which is
\bge
    \omega^2=k_\text{phys}(k_\text{phys}\pm 2\mu )+m^2-\FR{H^2}{4}.
\ede
The oscillation signal is possible only when $m>H/2$. Here again, we see that the particle production is most effective when $k_\text{phys}=\mu$ where $\omega$ is minimized. However, we note that the energy $\omega^2$ could become negative at this minimum if $\mu>m$. This is nothing but the usual condition that, in thermal equilibrium, the chemical potential of a boson system cannot be larger than its energy ($m$ in this case). When $\mu>m$, a tachyonic instability is generated at the minimum and this can lead to genuine exponential enhancement of particle production, which can sometimes be dangerous to the inflation background. The perturbation calculation also breaks down when $\mu$ is significantly larger than $m$, so we will not pursue this possibility further, but only restrict ourselves to the range where $\mu$ is at most comparable to $m$.

To see this point more quantitatively, we solve the above equations (\ref{Aeom}) and fine the following solutions,
\begin{align}
  &A_\pm = \FR{e^{\mp\pi\mu/2H}}{\sqrt{2k}} \text{W}_{\pm\ka,\nuv}(2\ii k\tau),
  &&A_3 = \FR{\sqrt{\pi}}{2}e^{\ii\pi\nuv/2}\sqrt{-\tau}\text{H}_{\nuv}^{(1)}(-k\tau),
\end{align}
The indices are $\ka\equiv \ii\mu/H$, $\nuv\equiv \sqrt{1/4-(m/H)^2}$. The normalization is determined by the canonical commutation relation $[A_i(\mb x),A_j'(\mb y)]=\ii\de_{ij}\de^{(3)}(\mb x-\mb y)$. The late-time $(\tau\to0)$ behavior of these modes are 
\begin{align}
  &A_\pm\simeq \FR{e^{-\ii \pi(1/4+\nuv/2)}}{\sqrt{2k}}\FR{e^{\mp \pi\mu/2H}\Gamma(-2\nuv)}{\Gamma(\frac{1}{2}\mp\ii\frac{\mu}{H}-\nuv)} (-2k\tau)^{\nuv+1/2}+(\nuv\to-\nuv),\\
  &A_3\simeq\FR{e^{-\ii\pi(1/2-\nuv/2)}}{\sqrt{2k}}\FR{\Gamma(-\nuv)}{\sqrt{\pi}}(-k\tau/2)^{\nuv+1/2}+(\nuv\to-\nuv).
\end{align}

  The chemical potential will bring exponential enhancement/suppression to the two transverse polarizations, but will not affect the ``apparent mass'' that shows up in the oscillation frequency and in the Boltzmann suppression factor. This means in particular that the enhancement from the chemical potential for a gauge boson is truly exponential. This can easily introduce gauge boson overproduction and thus huge back reaction to the inflation potential. Clearly the exponential enhancement will be significant when $\mu\gtrsim \max\{m,H\}$.
  However, we need to be careful about the size of the chemical potential in making approximations to estimate the NG.
A chemical potential $\mu\sim m$ will help to avoid the Boltzmann suppression even if  $m\gg H$. At the same time,   $\mu\gg m$ is not allowed due to the tachyonic instability discussed earlier in this section.   To estimate the NG, we can take the gauge boson propagator as $e^{-\pi(m-\mu)}$. However, one should always keep in mind that this no longer works when $\mu\gtrsim\max\{m,H\}$. We also note that, for typical $m\sim    H$, $\mu (\sim m) \sim  H$  can actually put a rather stringent bound on the cutoff scale $\Lambda$. One interesting possibility is that the gauge boson acquires a large mass if there is gauge symmetry breaking from a scalar field with a large VEV. Then one can introduce a large chemical potential to remove the Boltzmann suppression and this could potentially lead to a large signal.
 
On the other hand,  the longitudinal component is not affected by the chemical potential.    So we will not see the effect of chemical potential from the tree-level diagrams in 3-point function, because the gauge boson internal lines in tree-level 3-point function must be longitudinally polarized.\footnote{In inflation background there is no boost symmetry, so it is possible to draw a distinction between transverse and longitudinal polarizations without specifying the momentum.} Therefore the chemical potential appears at 1-loop. The 4-point function can nevertheless have transverse polarizations at the tree level. The chemical potential can then enhance the trispectrum greatly. This opens up a possibility to see CP violating signals in the tripspectrum. See \cite{Liu:2019fag}.

\subsection{Spin-2 particle}

A chemical potential can also be introduced for spin-2 particles, namely gravitons. When a graviton is massless, it can propagate to the end of inflation and thus we would expect to see tree-level signals of this chemical potential with graviton external legs, different from previous cases. For massive spin-2 particles, this chemical potential could also lead to enhanced particle production, similar to the case of a massive gauge boson.

For the massless graviton, we consider the following Lagrangian,
\begin{align}
    \ld = \FR{M_P^2}{2}\bigg[\sqrt{-g}R+\FR{1}{4\Lambda^3}\phi W_{\mu\nu\rh\si}\wt W^{\mu\nu\rh\si}\bigg],
\end{align}
where $W_{\mu\nu\rh\si}$ is the Weyl tensor. The equation of motion for the metric $g_{\mu\nu}$ is \cite{Jackiw:2003pm},
\bge
    G^{\mu\nu}+\FR{1}{2\Lambda^3\sqrt{-g}}\bigg[(D_\si\phi)(\ep^{\si\mu\al\be}D_\al R_\be^\nu+\ep^{\si\nu\al\be}D_\al R_\be^\mu)+(D_\si D_\tau \phi)(\wt R^{\tau\mu\si\nu}+\wt R^{\tau\nu\si\mu})\bigg]=0.
\ede
To derive the equation of motion for the tensor perturbation $h_{\mu\nu}$, we perturb the above equation of motion to the linear order, namely taking $g_{\mu\nu}=\bar g_{\mu\nu}+h_{\mu\nu}$, with $\bar g_{\mu\nu}$ the background metric which can be written as $\bar g_{\mu\nu}=a^2(\tau)\eta_{\mu\nu}$ in the conformal coordinates. The tensor perturbation satisfies the conditions $h_{0\mu}=0$, $h_{ii}=0$, $\pd_i h_{ij}=0$. We also choose $h_{ij}$ to have fixed 3-momentum $k_i=(0,0,k)$. Then its two independent polarizations can be taken to be $h_+=h_{11}=-h_{22}$ and $h_\times = h_{12}=h_{21}$. The equations for the two polarizations $(h_+,h_\times)$ are
\begin{align}
    &h_+''-\FR{2}{\tau}h_+'+k^2h_+=-\ii \nu_2 k \tau \bigg(h_\times''-\FR{4}{\tau}h_\times'+k^2h_\times\bigg),\\
    &h_\times''-\FR{2}{\tau}h_\times'+k^2h_\times=+\ii \nu_2    k \tau \bigg(h_+''-\FR{4}{\tau}h_+'+k^2h_+\bigg),
\end{align}
where the dimensionless parameter $\nu_2 \equiv    H\dot\phi_0/\Lambda^3$. Here we have taken $\dot\phi_0$ to be a constant and neglected terms involving $\ddot\phi_0$.
Define $h_{\pm}=h_+ \pm \ii h_\times$, we have the separated equations, 
\begin{align}
       h_{\pm}''-\FR{2}{\tau}\bigg(1\pm \FR{\nu_2 k\tau}{1\pm \nu_2 k\tau}\bigg)h_{\pm}'+k^2h_{\pm} =0.
\end{align}
This equation is not quite the same with the previous cases of spin-1/2 fermion and gauge boson, due to a more complicated first-order derivative term. However, we observe that, for any $\Lambda>\dot\phi_0^{1/2}$, the dimensionless parameter $\nu_2\ll 1$. Therefore, if we do not care about the modes deeply inside the horizon, it is possible to take $\nu_2 |k\tau|\ll 1$ as well. So we will just keep the leading dependence on $\nu_2$, and find the following solution for the mode,
\begin{align}
    h_{\pm}\sim\FR{\tau}{\sqrt{2k}}e^{\pm \nu_2 k\tau}\mathrm{W}_{\pm\ii\nu_2,3/2}(2\ii k\tau).
\end{align}
This mode function is pathological in the early time limit so we will not care about the overall normalization. But one can again see the effect of chemical potential $\zeta$ as enhancing one helicity state while suppressing the other. Since these modes approach constant in the late time limit $\tau\to 0$, the chemical potential could lead to an observable effect already at the tree level with graviton external legs, namely a circularly polarized tensor power spectrum. This effect has been known for a long time \cite{Lue:1998mq}. But there is also a similar effect for massive spin-2 particles as internal legs which could be interesting, too. We leave this possibility for a future study.

\section{Size of the Oscillatory Signal with Chemical Potential}
\label{sec_size_chempot}

Now we revisit the signal size with the presence of chemical potential. We will consider chemical potentials for scalars, fermions, and gauge bosons, respectively.  It would be interesting to consider massive spin-2 particles as well, we leave this to a future study.

As we showed in previous section, the presence of a chemical potential can sometimes enhance the signal in a way that is not captured in the naive estimate outline in Sec.~\ref{sec_size}. In this section we will include this additional enhancement in the estimate of signal size without detailed calculation.  We will be interested in the following set of diagrams.
\bge
\label{chempot1loop}
 \parbox{0.3\textwidth}{\includegraphics[width=0.3\textwidth]{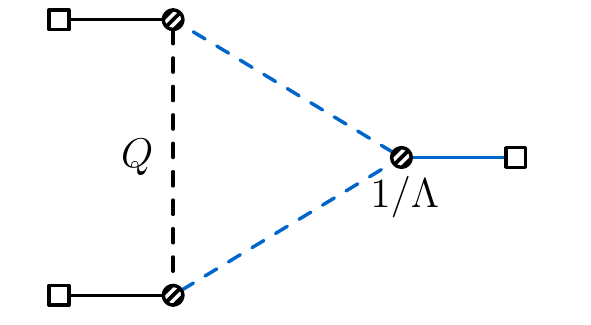}}
 \parbox{0.3\textwidth}{\includegraphics[width=0.3\textwidth]{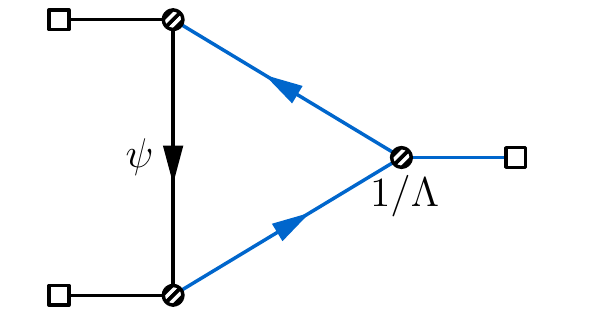}}
 \parbox{0.3\textwidth}{\includegraphics[width=0.3\textwidth]{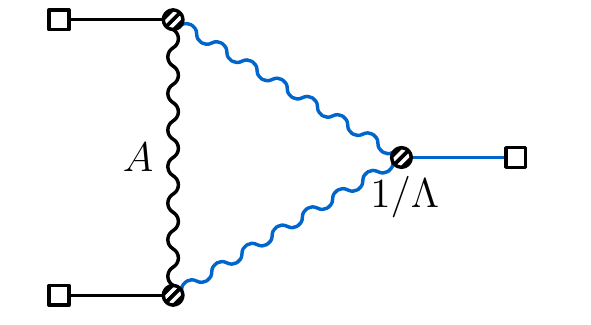}}
\ede
In each of these diagrams, we consider the minimal scenario where the inflaton-matter coupling is from the same dim-5 operator that gives chemical potential to the loop fields. Therefore each vertex will be associated with a factor of $H/\Lambda$. In Sec.~\ref{sec_size} we showed that a naive estimate of such diagrams would be the following if all internal propagators are taken to be $\order{1}$,
\bge
\label{fnlCPnaive}
  \fo\sim\FR{1}{2\pi P_\zeta^{1/2}}\FR{1}{16\pi^2}\Big(\FR{H}{\Lambda}\Big)^3\lesssim\FR{\sqrt{2\pi} P_\zeta^{1/4}}{16\pi^2}.
\ede
The final inequality follows from $\Lambda^2 \gtrsim \dot\phi_0$ as required by a valid EFT expansion in $\pd\phi/\Lambda^2$. However, in the presence of chemical potentials, the propagators cannot always be estimated as $\order{1}$. A more careful treatment would be necessary. In the following, we consider the three diagrams separately. 

\paragraph{Scalar.} In the case of a scalar field, there is no new enhancement of particle production as discussed above. On the other hand, in the most general case where the chemical potential and the mass terms break the $U(1)$ symmetry, namely
\begin{align}
  \Delta\ld=\sqrt{-g}\bigg[-\FR{1}{2}m_1^2Q_1^2-\FR{1}{2}m_2^2Q_2^2+\FR{c_1}{\Lambda}(\pd_\mu\phi)(\pd^\mu Q_1)Q_2+\FR{c_2}{\Lambda}(\pd_\mu\phi)(\pd^\mu Q_2)Q_1\bigg],
\end{align}
the presence of dim-5 operators would correct the mass of the scalars, shown in (\ref{masseigenvalue}), with two mass eigenvalues given by 
\begin{align} 
   m_\pm^2=\FR{1}{2}(m_1^2+m_2^2)+\FR{1}{4}(c_1-c_2)^2\FR{\dot\phi_0^2}{\Lambda^2}\pm\FR{1}{2}\sqrt{(m_1^2-m_2^2)^2+9H^2(c_1+c_2)^2\FR{\dot\phi_0^2}{\Lambda^2}}.
\end{align}
Therefore an improved estimate of the signal from the $Q$-loop is
\bge
  \fo \left(\parbox{0.25\textwidth}{\includegraphics[width=0.25\textwidth]{Fig_CP1loopQ}}\right)\sim \FR{1}{2\pi P_\zeta^{1/2}}\FR{1}{16\pi^2}\Big(\FR{H}{\Lambda}\Big)^3\FR{H}{m_-}e^{-2\pi m_-/H}.
\ede
Here we are assuming that $c_{1,2}\sim\order{1}$. The additional suppression factors are from the loop propagators. For the hard (black) line, we can approximate the propagator by its EFT limit $H/m_-$ when $m_->H$. With each of the soft (blue) lines, we should associate a Boltzmann suppression factor $e^{-\pi m_-}$. There are contributions from both mass eigenstates but we are only considering the one that is less suppressed, namely $m_-$. From this result, we can see that the signal is always tiny. One could tune the parameters such that $m_-\sim H$ while $\Lambda\sim\dot\phi_0^{1/2}$, and then the suppressions from the propagators disappear and the naive estimate (\ref{fnlCPnaive}) applies. However, this is still too small to be observed. 

\paragraph{Fermion.} We have shown that the operator $(\pd_\mu\phi)\psi^\dag\ob\si^\dag\psi/\Lambda$ can lead to enhancement of particle production. Now we use the result of Sec.~\ref{sec_mode} to estimate the middle diagram in (\ref{chempot1loop}). First, we observe that the vertex becomes unphysical when $m=0$, so one must associate each vertex with not only a cutoff scale $1/\Lambda$ but also a factor of $m$. 
For the hard (black) propagator in the loop, we can still approximate it by the EFT limit $1/ m$ when $m\gg H$. For each of the two soft loop lines, we need to include an exponential factor $e^{\pi\mu/H}$ for the particle production, and another exponential factor $e^{-\pi\sqrt{m^2+\mu^2}/H}$ for the Boltzmann suppression. In the most interesting parameter range $\mu\gg m\gg H$, the two factors combined into 
\bge
  \exp(\pi\mu)\exp\big(-\pi\sqrt{m^2+\mu^2}/H\big)\simeq \exp\big[-\pi m^2/(2H\mu)\big].
\ede
Therefore there is no obvious exponential enhancement/suppression when $\mu\gg m$. So a more precise estimate should also include power dependences on these parameters which is not obvious at the level of Feynman rules. One way to make progress is to note that the soft lines contribute to the signal only when they are on shell. Therefore the loop integral is actually counting the number of on-shell fermions that are produced through the non-adiabatic evolution. A measure of the non-adiabatic production is the parameter $-\dot\omega/\omega^2$ where $\omega$ is the physical energy of the fermion mode given by
\bge
  \omega^2=(k_\text{phys}\pm \mu)^2+m^2.
\ede
The non-adiabatic production is triggered when $\dot\omega/\omega^2$ becomes large. In the inflation background with $k_\text{phys}=k/a$, one can see that $\dot\omega/\omega^2$ remains tiny at both early and late times, and it develops a peak when $k_\text{phys}\simeq \mu$ with a width $\Delta k_\text{phys}\simeq m$. Therefore, the total number of on-shell fermions contributing to the middle diagrams are from a momentum shell with radius $\mu$ and width $m$. So the amplitude of the signal should be proportional to the volume of this momentum shell, which is $4\pi m\mu^2$. 

Taking all ingredients into account, we can now estimate the fermion diagram as
\begin{align}
  \fo\left( \parbox{0.25\textwidth}{\includegraphics[width=0.25\textwidth]{Fig_1loopFermion}}\right)\sim \FR{1}{2\pi P_\zeta^{1/2}}\FR{1}{16\pi^2}\Big(\FR{m}{\Lambda}\Big)^3\FR{H}{m}\cdot 4\pi m\mu^2\cdot e^{-\pi m^2/(\mu H)}.
\end{align}
Using the relation $\mu=\dot\phi_0/\Lambda$, this can be rewritten as
\bge
\label{fnlfermion}
  \fo\sim \FR{P_\zeta}{2}\Big(\FR{m}{H}\Big)^3\Big(\FR{\mu}{H}\Big)^5e^{-\pi m^2/(\mu H)}.
\ede
This agrees parametrically with the result in \cite{Chen:2018xck}, and it can be as large as $\order{10}$. And there can be additional enhancement when there are multiple degrees of freedom contributing the loop. For instance, in the case of a SM quark running in the loop, we have an additional factor of $2\times 3$ with $2$ from Dirac fermion being two copies of Weyl fermions and $3$ from the color factor. 

\paragraph{Gauge boson.} Finally we consider briefly the case of gauge boson. We first note that the constraint that chemical potential $\mu=\dot\phi_0/\Lambda$ cannot be significantly larger than mass $m$ can put a stringent bound on the cutoff scale $\Lambda$ for $m\simeq H$. In this case $\Lambda>\dot\phi_0/H\simeq 3600H$, and the right diagram in (\ref{chempot1loop}) will be tiny. Therefore one need to consider either a new coupling that couples the inflaton to the gauge boson, or a large mass being generated to the gauge boson via symmetry breaking. These will not be as economic as the fermions case but could still be interesting. We leave a detailed study of this scenario for a future work.

\section{Results and Discussions}
\label{sec_discussions}

The primordial NG could carry signals of heavy particle production during inflation and thus provide a unique window for probing new physics at very high scales. However, the size of the signal is very sensitive to model details. On the observational side, LSS survey in the near future could improve the measurement of NG by roughly one order of magnitude, with an additional two orders of magnitude attainable by more futuristic 21cm tomography. It is therefore important to understand which type of models could possibly generate signals large enough to be observed in the near future. 

In this paper we considered this problem in a minimal scenario of inflation. By minimal we mean that the inflaton generates all primordial fluctuations which are scale invariant and respect all spacetime symmetries, up to slow-roll corrections. In addition, we assume that the inflaton couple to matter fields through EFT couplings with a single cutoff scale and all dimensionless couplings being $\order{1}$. We also avoid any accidental cancelation or tuning of parameters at the tree level. Imposing all these conditions lead to an almost no-go result, with a single exception of the dimension-5 couplings that generate a parity-odd chemical potential for the matter fields.

We then explained how the parity-odd chemical potentials enhance the particle production and increase the signal, by both intuitive arguments and explicit calculating the mode functions.

To put our results into context, we show in Fig.~\ref{Fig_signal} the signals from chemical potential channels, together with some other known scenarios that could generate large signals. In this plot, the horizontal direction shows the oscillation frequency $|\nu|$ of the signal, while the vertical direction shows the amplitude, both of which are observable at least in principle. 

The shaded region of a singlet Majorana fermion (for instance, a heavy right-handed neutrino) come from a more careful calculation in \cite{Chen:2018xck,Hook:2019zxa}, which is larger than our simple estimate in (\ref{fnlfermion}) by a factor of $\sim 2$.  Another factor of 6 is included for SM quarks other than the top. These signals are possible if the SM Higgs develop a new minimum during the inflation. In each of the two shaded regions, we vary the cutoff scale from $\Lambda^2=2\dot\phi_0$ (the right edge) towards larger values (moving towards the lower-left direction), and for each $\Lambda$, we vary the mass $m$ of the fermion also. We show a particular example of neutrino signal with fixed cutoff scale $\Lambda=2\dot\phi_0^{1/2}$ and changing mass in dark blue curve. 

The top quark is special in that the top condensate could by itself generate a new VEV for the Higgs, and thus the amplitude is uniquely related to the oscillation frequency.

In the same plot we also include, for comparison, signals from tree-level exchange of $Z$ boson in a cosmological Higgs collider scenario, taken from Eq. (69) of \cite{Lu:2019tjj}; the tree-level Higgs exchange from a curvaton scenario, taken from Eq. (5.16) of \cite{Kumar:2019ebj}; a tree-level KK graviton exchange in an extra dimension GUT model, taken from Eq. (6.2) of \cite{Kumar:2018jxz}; and finally, the signals from the quasi single field inflation, discussed in Sec.~\ref{sec_size}, and the signals are taken from results in \cite{Chen:2017ryl}. The dotted segments for $\lam=10^{-6}$ and $10^{-7}$ indicate that we are extrapolating the result of our perturbative estimate into the strong coupling regime where the two point mixing is greater than $\max\{H,m\}$ \cite{An:2017hlx,Iyer:2017qzw}.  

We note that the chemical potential scenario is special in that it populates the parameter space with large signal \emph{and} large frequency. In general, an unsuppressed signal with large frequency implies a new mechanism of particle production above the Hubble scale. 

In the same figure we also include, very roughly, the current constraint from Planck \cite{Akrami:2019izv}, and also the future reach of LSS survey and 21cm tomography. All these constraints should be shape- and template-dependent. In particular, we note that the current Planck constraint on equilateral shape does not extend to regions with large oscillation frequency, and we acknowledge this fact by not extending the Planck constraint to large $|\nu|$ region, showing that a large signal  with very large frequency is not yet ruled out. On the other hand, we are simply drawing horizontal lines for LSS and 21 cm observations, assuming that appropriate templates are adopted even for large frequency signals. 

In this paper, we have considered minimal scenarios in the EFT framework, with simple naturalness assumptions. There are several obvious future directions to pursue. Several interesting scenarios mentioned in the paper, including extended models with gauge bosons and massive spin-2 particles, merit a fuller exploration.   It would also be interesting to consider well motivated UV models which give rise to the chemical potential coupling highlighted here. More sophisticated models can also potentially alleviate the apparent fine-tuning in some of the other cases and lead naturally to large NG signals. Our results also motivate further work in this direction.

\paragraph{Acknowledgment.} We thank Xingang Chen, Anson Hook, Junwu Huang, Soubhik Kumar, Raman Sundrum, and Yi Wang for useful discussions. LTW is supported by the DOE grant DE-SC0013642.

\providecommand{\href}[2]{#2}\begingroup\raggedright\endgroup

\end{document}